\documentclass[conference]{IEEEtran}
\ifCLASSINFOpdf
   \usepackage[pdftex]{graphicx}
   \graphicspath{{../pdf/}{../jpeg/}}
   \DeclareGraphicsExtensions{.pdf,.jpeg,.png}
\else
   \usepackage[dvips]{graphicx}
   \graphicspath{{../eps/}}
   \DeclareGraphicsExtensions{.eps}
\fi

\ifCLASSOPTIONcompsoc
  \usepackage[nocompress]{cite}
\else
  \usepackage{cite}
\fi

\begin{document}
\title{Predicting missing links via correlation between nodes}

 %author names and affiliations
\author{\IEEEauthorblockN{Hao Liao}
\IEEEauthorblockA{Department of Physics\\
University of Fribourg\\
Fribourg, Switzerland CH-1700\\
}
\and
\IEEEauthorblockN{An Zeng}
\IEEEauthorblockA{Department of Physics\\
University of Fribourg\\
Fribourg, Switzerland CH-1700\\
Email: an.zeng@unifr.ch}
\and
\IEEEauthorblockN{Yi-Cheng Zhang}
\IEEEauthorblockA{Department of Physics\\
University of Fribourg\\
Fribourg, Switzerland CH-1700\\
Web Sciences Center,\\
School of Computer Science and Engineering,\\
University of Electronic Science and Technology of China,\\
Chengdu 611731, People’s Republic of China\\
}}

\maketitle

\begin{abstract}
%\boldmath
As a fundamental problem in many different fields, link prediction aims to estimate the likelihood of an existing link between two nodes based on the observed information. Since this problem is related to many applications ranging from uncovering missing data to predicting the evolution of networks, link prediction has been intensively investigated recently and many methods have been proposed so far. The essential challenge of link prediction is to estimate the similarity between nodes. Most of the existing methods are based on the common neighbor index and its variants. In this paper, we propose to calculate the similarity between nodes by the correlation coefficient. This method is found to be very effective when applied to calculate similarity based on high order paths. We finally fuse the correlation-based method with the resource allocation method, and find that the combined method can substantially outperform the existing methods, especially in sparse networks.
\end{abstract}


% \fi
\IEEEpeerreviewmaketitle

\section{Introduction}
The object of many scientific researches is prediction. For instance, understanding the mechanism of epidemic spreading can help us to predict the future coverage of a certain virus~\cite{science13}, the mechanistic model for the citation dynamics of individual papers can be applied to predict the future evolution of scientific publications~\cite{DW2013}. While mathematical models and prediction techniques are sufficiently mature for some systems, reliable prediction approaches are still unavailable in most systems. One may identify their chaotic nature to be the major difficulty, yet the lack of understanding of the underlying principles may indeed be the main obstacle. Besides the prediction of the collective behavior, the prediction in microscopic level, such as the well-known link prediction challenge in complex networks, has also attracted a lot of attention.

Link prediction is a very important problem that aims at estimating the likelihood of the existence of a link between two nodes ~\cite{LNK07,LK03}. Solving this problem cannot only help us complete the missing data in biological networks such as the protein-protein interaction networks and metabolic networks\cite{Nature1,Nature2}, but also enable us to predict the evolution of social networks \cite{AA03,WPSGB11,icdma13}. In fact, link prediction is also closely connected to other problems such as recommendation \cite{ZLZZ09} and spurious links detection \cite{An12}. A sound link prediction method will help to design more efficient recommendation algorithm to filter out irrelevant information for online users\cite{icdms13}. Moreover, the link prediction method can be also applied to analyze the reliability of existing links and accordingly identify some noisy connections from networks. The progress in this field will largely push forward the researches in other fields. Accordingly, the problem of missing link prediction has been intensively studied by researchers from different backgrounds and many methods applied to different fields have been proposed \cite{LJN2010,LSTD10,cikm12,LT09}. For a review, see ref.\cite{LuZ2010}.

The basic assumption for link prediction is that two nodes are more likely to have a link if they are similar to each other. Therefore, the essential problem for link prediction is how to calculate the similarity between nodes accurately. One of the most straight-forward method is called common neighbor which measures the similarity between two individuals by directly counting the number of common neighboring nodes \cite{CN1971}. However, this method has serious shortcomings as it strongly favors the large degree nodes. To solve this problem, many variants, such as the Jaccard index \cite{Jaccard1901} and Salton index \cite{Salton1975}, have been proposed to remove this tendency. In addition, some other methods including Katz index \cite{Katz1953}, simrank \cite{GJ2002},hierarchical random graph \cite{Nature2} and stochastic block model \cite{BM2011,icdml13}are also very effective in estimating nodes' similarity. However, these methods are based on global algorithms that can be prohibitive to use for large-scale systems.

In this paper, we argue that the similarity between nodes can be calculated based on another completely different type of method, called correlation. In broad definition, correlation refers to any class of statistical relationships involving dependence between two or more random variables. In our case, we actually refer it to the Pearson correlation \cite{LIN1989}between nodes' attribute vectors which can come from the adjacency matrix or higher order of that. In link prediction, one of the biggest challenge is the data sparsity. It means that a lot of data at hand is too sparse to extract valuable similarity information from the simple common neighbor method or its variants. One possible solution has been discussed in ref.\cite{LP2009} in which longer paths (i.e. paths with length larger than 2) are applied to measure nodes' similarity. However, when it comes to such high order information much noise will be included so that the similarity matrix is indeed denser but the similarities are not satisfactorily accurate, which leads to a poor outcome of predicted links. In our simulation, we find that the correlation-based method is very effective when applied to calculate similarity based on high order paths. We finally fuse the new method with the resource allocation method \cite{RA2009}, and find that the combined method can substantially outperform the existing methods, especially in sparse networks.

\section{Related works}
To begin our analysis, we first briefly describe the link prediction problem and review some representative methods. Considering an unweighted undirected simple network $G(V,E)$, $V$ is the set of nodes and $E$ is the set of links. The multiple links and self-connections are not allowed. For each pair of nodes $x$, $y$ belonging to $V$, we calculate a score $s_{xy}$ which measures the likelihood for node $x$ and $y$ to have a link between them.
Since $G$ is undirected, the score is supposed to be symmetry, i.e. $s_{xy} = s_{yx}$. All the nonexistent links are sorted
in decreasing order according to their $s$ scores, and the links on the top are most likely to exist. There are many different ways to calculate $s_{xy}$ score and the most common and straightforward way is to calculate the similarity between node $x$ and $y$.

Generally speaking, two nodes are considered to be similar if they have some common important features in topology. An review paper on these similarity indices in ref.\cite{LuZ2010}. In this paper,we compare the prediction accuracies of four typical similarity indices: Common neighbor (CN), Resource allocation (RA), Jaccard, Local path. Their definitions and relevant motivations are introduced as follows:

(i) \emph{Common Neighbor} (\textsf{CN}).
Two nodes $x$ and $y$ are more likely to form a link if they have many common neighbors. Let $\Gamma(x)$ denote the set of neighbors of node $x$,
the simplest measure of the neighborhood overlap can be the directly calculated as:
\begin{equation}
  \label{eq:cn}
  s^{\rm CN}_{xy} = |\Gamma(x)\cap\Gamma(y)|
\end{equation}
which is the actual aggregation method used by most websites. However, the drawback of CN is that it is in favor of the nodes with large degree. It is obvious that $s_{xy} =(A^2)_{xy}$, where $A$ is the adjacency matrix, in which $A_{xy}=1$ if $x$ and $y$ are directly connected and $A_{xy}=0$  otherwise. Note that $(A^2)_{xy}$ is also the number of different paths with length $2$ connecting $x$ and $y$. Newman \cite{LuZ2010} used this quantity in the study of collaboration networks, showing the correlation between the number of common neighbors and the probability that two scientists will collaborate in the future.Therefore, we here select CN as the representative of all CN-based measures. Although CN consumes little time and performs relatively good among many local indices, due to the insufficient information, its accuracy cannot catch up with the measures based on global information. One typical example is the Katz index\cite{Katz1953}.

(ii) \emph{Jaccard coefficient} (\textsf{Jaccard}).
This index was proposed by Jaccard over a hundred years ago, The algorithm is a traditional similarity measurement in the literature. It is defined as
\begin{equation}
  \label{eq:jaccard}
  s^{\rm Jaccard}_{xy} = \frac{|\Gamma(x)\cap\Gamma(y)|}{|\Gamma(x)\cup\Gamma(y)|}.
\end{equation}
The motivation of this index is that the pure common neighbor favors a lot for the large degree nodes: it is easier for large degree nodes to form common neighbors with other nodes. The denominator can remove the tendency for high degree nodes to have high similarity with other nodes. Note that, there are many other ways to remove the tendency of CN to large degree nodes, such as cosine index, Sorensen index, Hub promoted index and so on~\cite{LuZ2010}.

(iii) \emph{Resource allocation} (\textsf{RA}).
This index is motivated by the resource allocation dynamics on complex system. Considering a pair of nodes, x and y, which are not directly connected, it assumes that the node $x$ needs send some resource to $y$, with their common neighbors playing the role of transmitters.Each transmitter has a unit of resource and will equally distribute it to all its neighbors. In the simplest case, we assume that each transmitter has a unit of resource, and will equally distribute it to all its neighbors. The similarity between $x$ and $y$ is define as the amount of resource $y$ received from $x$:
\begin{equation}
 \label{eq:RA}
 s^{\rm RA}_{xy} = \sum_{z\in\Gamma(x)\cap\Gamma(y)}\frac{1}{k_{z}}.
 \end{equation}
It is obvious that this measure is symmetric, namely $s_{xy}=x_{yx}$. A similar similarity index is called Adamic-Adar (AA) Index which simply replaces $k_{z}$ in the above equation by $logk_z$. Although resulting from different motivations, the AA index and RA index have very similar forms. Indeed, they both depress the contribution of the high-degree common neighbors. AA index takes the form $(logk_z)^{-1}$ while RA index takes the form $k_z^{-1}$. The different is insignificant when the degree, $k_z$ is small, while it is considerable when $k_z$ is large. So RA index punished the high degree common neighbors more heavily than AA. Previous study showed that RA performs the best among all the common-neighbor-based methods in the USAir network, NetScience network, Power Grid network, etc.

(iv) \emph{Local Path index} (\textsf{LP}).
This index was introduced by ref\cite{LP2009}. This index takes local paths into consideration, with wider horizon than CN.It is given by
\begin{equation}
  \label{eq:lp}
  S^{\rm LP}_{xy}=A^2+\epsilon*A^3
\end{equation}
where $\epsilon$ is chosen as a free parameter and $A$ is the adjacency matrix of the network. Clearly, this measure degenerates to CN when $\epsilon=0$. And if $x$ and $y$ are not directly connected, $(A^3)_{xy}$ is equal to the number of different paths with length $3$ connecting $x$ and $y$. This index can be extended to account for higher-order paths, as
$s^{LP(m)}=A^2+\epsilon(A^3)+\epsilon^2(A^4)+...+\epsilon^{m-2}(A^m)$,
where $m>2$ is maximal order. When $n$ equals to the number of nodes in the network, LP is equivalent to the well-known Katz index~\cite{Katz1953} which takes all paths into account in the network. The computational complexity of this index in an uncorrelated network is $O(N(k^m))$,which grows fast with the increasing of n and will exceed the complexity for calculating the Katz index (around to $O(N^3)$) for large $m$. Experimental results show that the optimal $m$ is positively correlated with the average shortest distance of the network.

\section{Similarity based on correlation between nodes}
The above methods, though effective in link prediction, all measures the similarity between nodes based on the common neighbor information. In this paper, we propose to calculate node similarities based on the correlation between nodes. In fact, similar idea has been applied to design ranking algorithm for online users' reputation~\cite{LA2014}. Given a vector $\textbf{v}_x$ ($\textbf{v}_y$) describing the feature of a node $x$ ($y$), we calculate the similarity between these two nodes based on the Pearson correlation coefficient of $v_x$ and $v_y$. Mathematically, it reads
\begin{equation}
  \label{eq:corr}
  s^{\rm Corr}_{xy}= \frac{1}{N}\sum_{l=1}^{N}\frac{{{(v_{xl}-\bar{v}_x)}{(v_{yl}-\bar{v}_y)}}}{\sigma_{v_x}\sigma_{v_y}}
\end{equation}
where $\bar{v}_x$ and $\sigma_{v_x}$ are respectively the mean and standard deviation of vector $\textbf{v}_x$. As discussed above, $\textbf{v}_x$ should be a attribute vector for node $x$. One simple way would be directly set as $v_{xl}=A_{xl}$. In this paper, we go beyond the adjacency matrix and take $A^m$ into consideration ($m$ can be larger than $1$), so that we set $\textbf{v}_x$ as the corresponding column of the $A^m$.

\begin{figure}[t!]
\centering
\includegraphics[width=9cm]{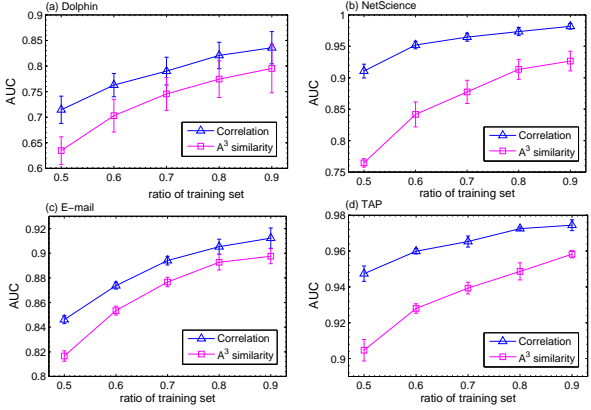}
\caption{(colour online) AUC of the $s^{\rm Corr}_{xy}$ and $A^3$ methods as a function of $1-p$ (the fraction of links in the training set) in four real networks. The error bars are obtained based on 10 independent realizations.}
\label{fig:scale}
\end{figure}

\begin{figure}[t!]
\centering
\includegraphics[width=9cm]{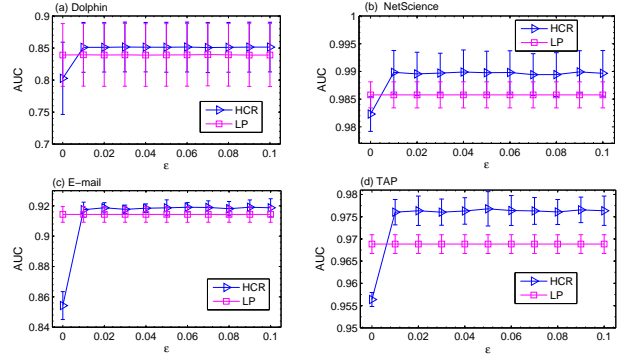}
\caption{
  (colour online) The dependence of the AUC of the HCR method on $\epsilon$ in four real networks. The results of the LP method are shown for comparison. In the LP method, the parameter is chosen as $\epsilon=0.01$ which is shown to be the optimal parameter for this method according to ref.\cite{LP2009}. The error bars in this figure are obtained based on 10 independent realizations.
}
\label{fig:scale}
\end{figure}

\section{Data and Metrics}
\begin{table*}
\caption{Structural properties of the different real networks. Structural properties include network size ($N$), edge number ($E$), degree Heterogeneity ($H=\langle k^2\rangle/\langle k\rangle^2$), degree assortativity ($r$)\cite{netcoauthor_word}, clustering coefficient ($\langle C \rangle$)\cite{PRL2002} and average shortest path length ($\langle d \rangle$). }
\label{tab1}
\begin{center}
\begin{tabular}{p{1.6cm} p{1cm} p{1cm} p{1cm} p{1cm} p{1cm} p{1cm} p{1cm} p{1cm} p{1cm}}
\hline
\hline
Network &$N$ &$E$ &$H$ &$r$ &$\langle C \rangle$ &$\langle d \rangle$ \\
\hline
Dolphin            &62 &159 &1.327 &-0.044 &0.259  &3.357\\
Jazz               &198 &2742 &1.395 &0.020 &0.618 &2.235 \\
C.elegans          &297 &2148 &1.801 &-0.163 &0.292 &2.455\\
USAIR              &332 &2126 &3.464 &-0.208 &0.749 &2.46\\
Netscience         &379 &914 &1.663 &-0.082 &0.741  &6.042\\
Email              &1133 &5451 &1.942 &0.078 &0.220 &3.606\\
TAP                &1373 &6833 &1.644 &0.579 &0.529 &5.224\\
PowerGrid          &4941 &6594 &1.450 &0.004 &0.080 &18.989\\
HEP                &5835 &13815 &1.926 &0.185 &0.506 &7.026\\
\hline
\hline
\end{tabular}
\vspace*{0.0cm}
\end{center}
\end{table*}

To evaluate the effectiveness of the above methods, we consider nine empirical networks including both social networks and nonsocial networks:
(i)Dolphin: a dolphin friendship network, which is an undirected social network of frequent associations between 62 dolphins in a community living off Doubtful Sound, New Zealand.\cite{dolphins}. (ii)Jazz: a music collaboration network obtained from The Red Hot Jazz Archive digital database. Here it includes 198 bands that performed between 1912 and 1940, with most of the bands in the $1920$ to $1940$\cite{jazz}. (iii)C.elegans: the neural network of the nematode worm C.elegans, in which edge joins two neurons if they are connected, by either a synapse or a gap junction \cite{Power}. (iv)USAIR: the US air transportation network \cite{USAir}.which contains 332 airports and 2126 airlines. (v) Netscience: a coauthorship network between scientists who are publishing on the topic of network science \cite{netcoauthor_word}.This network contains 1589 scientists, and 128 of whom are isolated. Here we do not consider those isolated nodes. Actually, it is consisted of 268 connected components, and the size of the largest connected component is only 379. The connectivity of NS is not good.(vi) Email: an email communication network~\cite{email}. (vii) TAP: a yeast protein binding network generated by tandem affinity purification experiments \cite{TAP}. (viii) Power Grid: the electrical power grid of western US \cite{Power},with nodes representing generators, transformers and substations, and links corresponding to the high-voltage transmission lines between them. This network contains 4941 nodes, and they are well connected. (ix) HEP: a collaboration network of high energy physicists, which contains the collaboration network of scientists posting preprints on the high-energy theory archive at www.arxiv.org from 1995 to 1999\cite{netcoauthor_word}. We only take into account of the giant component of these networks. This is because for a pair of nodes located in two disconnected components, their $s_{xy}$ score will be zero according to CN and its variant. Table 1 shows the basic statistics of all the giant components of those networks.

For each of the real network, the observed links $M$ are randomly divided into two parts: the training set $M^T$, is treated
as known information, while probe set, $M^p$, is used for verifying the prediction accuracy and no information in which is permitted
to be used for prediction. $M^T$ plus $M^p$ is the whole data set. Note that, each time before moving a link to the probe set we first check if this removal will make the training network disconnected. Usually, the training set contains $90\%$ of the links and the probe set consists of $10\%$ links. In this paper, we employ a standard metric, area under the receiver operating characteristic curve (AUC)\cite{AUC1984} to measure the accuracy of the prediction. AUC can be interpreted as the probability that a randomly chosen missing link from $M^p$ is given a higher score than a randomly chosen nonexistent link. In practice, we usually calculate the score of each non-observed link instead of giving the ordered list since latter task is more time-consuming. Then, AUC requires $n$ times of independent comparisons. at each time we randomly choose a missing link and nonexistent link to compare their scores. After the comparison,we record there are $n_1$ times the missing link having a higher score, and $n_2$ times they have the same score.The final AUC is calculated as $AUC=(n_1+0.5*n_2)/n$. If all the scores are given by an independent and identical distribution, then $AUC$ should be around $0.5$. A higher AUC is corresponding to a more accurate prediction.

\section{Results}
\begin{table*}[t!]
\caption{Comparison of different algorithms' accuracy quantified by AUC or each real network considered. The training set contains $90\%$ of the known links. Each number is obtained by averaging over $10$ implementations with independently random divisions of training set and probe set. We set the parameters $\epsilon = 10^{-2}$ in LP and $\epsilon = 10^{-2}$ in HCR. The highest accuracy in each line is emphasized by boldface.}
\label{tab1}
\begin{center}
\begin{tabular}{p{1.6cm} p{1cm} p{1cm} p{1cm} p{1cm} p{1cm} p{1cm} p{1cm} p{1cm} p{1cm}}
\hline
\hline
Network & $CN$ & $RA$ & $Jaccard$ & $LP$ &$HCR$  \\
\hline
Dolphin           &0.803 &0.806 &0.802 &0.829 &\textbf{0.846}\\
Jazz               &0.955 &0.971 &0.958 &0.947 &\textbf{0.973} \\
C. elegans         &0.846 &0.867 &0.811 &0.866 &\textbf{0.881} \\
USAIR              &0.954 &0.972&0.915 &0.952 &\textbf{0.974}\\
Netsci             &0.981 &0.985 &0.981 &0.989 &\textbf{0.992} \\
Email              &0.858 &0.859 &0.856 &0.919 &\textbf{0.922} \\
TAP                &0.954 &0.954 &0.956 &0.967 &\textbf{0.977} \\
PowerGrid          &0.624 &0.624 &0.624 &0.689 &\textbf{0.767}\\
HEP                &0.941 &0.943 &0.942 &0.961 &\textbf{0.965}\\
\hline
\hline
\end{tabular}
\vspace*{0.0cm}
\end{center}
\end{table*}
We first compare the performance of our method in four representative data sets: Dolphin, Netscience, Email and TAP. The detailed values on other data sets will be reported in table 2. In fact, we observe in our simulation that $s^{\rm Corr}_{xy}$ itself cannot outperform the traditional similarity measure such as CN, Jaccard, RA and LP in link prediction. However, we find that $s^{\rm Corr}_{xy}$ work well in extracting the node similarity information from high order paths (i.e. paths with length larger than $2$). In order to show this, we set $m=2$ in $s^{\rm Corr}_{xy}$ and compare it with $s=A^3$ which directly uses the number of paths with length $3$ to measure the similarity between nodes. Note that $A^3$ method has already been applied to combine with the common neighbor index in the LP method to solve the data sparsity problem.

We investigate the effect of data sparsity on the $s^{\rm Corr}_{xy}$ and $A^3$ method. To this end, we move fraction $p$ of all links to the probe set and use the remaining $1-p$ links as the training set. A larger $p$ is corresponding to a more sparser known information of the real network. The AUC results of both methods under different $p$ are presented in Fig. 1. One can see that in all networks considered, AUC of the $s^{\rm Corr}_{xy}$ is significantly higher than that of $A^3$. Interestingly, the advantage of $s^{\rm Corr}_{xy}$ to $A^3$ becomes generally larger when the fraction of links in the training set is smaller. These results are actually very important since the high order paths are usually applied to solve the data sparsity problem. A better use of the high order paths information can solve the data sparsity problem more effectively.

Inspired by the results above, we propose to combine the $s^{\rm Corr}_{xy}$ method and one of the traditional similarity method to achieve higher accuracy in link prediction. As the RA method is one of the most efficient ones among the variants of the CN method, we adopt it to design the new method. As the new method is a Hybrid of the Correlation method and the Resource allocation method, we refer it as HCR method in this paper. The formula of the HCR method reads
\begin{equation}
 s^{\rm HCR}_{xy}=s^{\rm RA}_{xy}+\epsilon*s^{\rm Corr}_{xy}(m=2),
\end{equation}
where $\epsilon$ is a tunable parameter. In fact, $s^{\rm RA}_{xy}$ already enjoys a high prediction accuracy in dense data and $s^{\rm Corr}_{xy}$ can accurately predict missing links with very sparse information. Therefore, the HCR method is a very general link prediction method which is supposed to work both in dense and sparse networks.

To validate the HCR method, we study the dependence of its AUC on $\epsilon$ in four real networks in Fig. 2. One can see that the prediction results are substantially improved once $\epsilon$ is larger than $0$, which indicates that the $s^{\rm Corr}_{xy}$ method can indeed improve the $s^{\rm RA}_{xy}$ method (corresponding to $\epsilon$). Moreover, as the LP method was proposed to solve the data sparsity problem as well, we compare the HCR method with it in Fig. 2. One can see that, when $\epsilon>0$ HCR method can outperform the LP method, indicating the data sparsity problem is better addressed in the HCR method. This result is actually reasonable, as we already observe above that $s^{\rm Corr}_{xy}$ can outperform $A^3$ in sparse networks.

The results on the other networks are reported in table 2. One can see as well that the HCR method generally have higher AUC than other methods in all the networks considered. Among these networks, power grid is a very sparse network. The similarity indices based on local information, such as the CN, RA and Jaccard, are all with low AUC. Different normalization to CN in this case cannot make much difference. However, once the semi-local information are taken into account, the LP method can significantly improve the AUC by nearly $10\%$. Interestingly, the HCR method performs even better than LP and improve the AUC by more than $23\%$. Similar phenomenon can be seen in Email network as well. On the other hand, when the network is dense, such as the Jazz and USAir network, the LP method cannot outperform the CN method as the information from high order paths in this case is too noisy. In contrast, the HCR method have higher AUC than CN and other method, indicating that HCR can make better use of the high order paths information than LP.

In link prediction, it is generally difficult to predict the missing
link of the nodes with small degree. This is known as the "cold-start`` problem~\cite{LuZ2010}. In the literature, it has already been shown
that the item cold-start problem can be well addressed by changing the denominator in the CN method~\cite{coldstart}. More specifically,
the prediction accuracy for small degree nodes can be largely improved when larger score is given to the node pairs with small degree. However, in sparse networks the cold-start problem can not be effectively solved in this way. In other words, the AUC cannot be substantially increased by just changing the denominator in the CN method.

In fact, the essential difficulty for the cold-start problem is that the available information for the small degree nodes are too limited
for the algorithms to accurately predict their missing links. The LP and HCR can address the cold-start problem by incorporate more information from high order paths. In order to show this, we pick the nodes with degree smaller than $k$, and report the prediction accuracy (AUC) of the probe set links between them in Fig. 3. We compare CN, LP and HCR methods in Fig. 3. As expected, one can see that AUC generally increases with $k$, indicating that the links connecting small degree nodes are indeed more difficult to be correctly predicted. Moreover, it is clearly that the LP method can indeed result in a higher AUC than CN. The HCR method can significantly improve the AUC of small degree nodes. Therefore, we conclude that HCR is more effective in solving cold-start problem than the LP method. In fig. 3, we consider 4 real networks. Even though the results are generally the same, we observe that the advantage of HCR method is bigger in sparser networks.

\begin{figure}[t!]
\centering
\includegraphics[width=9cm]{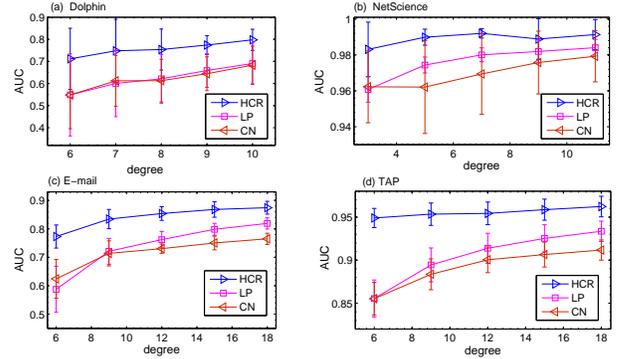}
\caption{(colour online) The AUC of the probe set links connecting nodes with degree sum smaller than $k$ when different link prediction algorithms are applied. In the LP method, the parameter is chosen as $\epsilon=0.01$. In HCR methdo, the parameter is chosen as $\epsilon=0.01$. The error bars are obtained based on 10 independent realizations.
}
\label{fig:scale}
\end{figure}

In principle, one can extend the current HCR method to deal with even higher order paths, and the modified HCR method reads
\begin{equation}
 s^{\rm HCR}_{xy}=s^{\rm RA}_{xy}+\epsilon\sum_ms^{\rm Corr}_{xy}(m).
\end{equation}
The results show that AUC can be slightly improved with higher order paths. We also consider an extension of the LP method,
\begin{equation}
 %s^{\rm LP}_{xy}= A^2+\epsilon\sum_m A^m.
  s^{\rm LP}_{xy}= A^2+\sum_m \epsilon^mA^m.
\end{equation}
However, the AUC of LP decreases when $m>3$. These results evidently supports again that HCR is more effective than LP in extracting similarity information, especially when the original information contains noise.

\section{Conclusion and discussion}
In this paper, we employ the Pearson correlation coefficient to measure the similarity between nodes, and accordingly apply it to predicting the future links. We find that though the correlation method cannot outperform the common neighbor and its variants in link prediction, the correlation method actually very efficient in extracting the similarity information from the high order path information. This is because the Pearson correlation coefficient is generally more robust to noise than the traditional index based on common neighbor. We further combine the correlation method and the resource allocation method, and find that this method can outperform the existing link prediction methods, especially when the available information from the observed network is little. We compare the new method with one existing method that intended to solve the data sparsity problem, and the results show that our method have higher accuracy.

Many issues remain still open. Our work implies that the Pearson correlation coefficient is more resistent to noisy information than the other methods. An interesting extension would be to investigate the link prediction problem in the noisy environment, i.e. the observed network containing some noisy links. One can compare the correlation-based method and the other method, and systematically study the robustness of these methods to noise. The Pearson correlation opens a new direction for measuring the similarity between nodes. In fact, there are some other correlation coefficients such as Spearman\cite{Rank1904} and Kendall's tau\cite{Tau1938} coefficient. A detailed study of their performance in link prediction would be another interesting extension.

Our results also show that the high order paths in networks also contain some valuable information to characterize node similarity. This information is especially important for sparse networks. Similar study has already been conducted in recommender systems where the semi-local diffusion is found to be able to significantly improve the recommendation accuracy~\cite{Plosone1}. However, if such information is not used properly, too much noise will be involved and may jeopardize the predict accuracy~\cite{NJP1}. Therefore, the link prediction method that is tolerant of noise is very important. In this paper, we present a possible method to solve this problem. There are some other possible ways for this problem, such as only taking into account the salient high order paths. Related methods ask for investigation in the future.

\section*{Acknowledgment}
This work was partially supported by the EU FP7 Grant 611272 (project GROWTHCOM) and by the Swiss National Science Foundation (grant no.~200020-143272).

\end{document}